\documentclass{article}
\usepackage{arxiv}
\usepackage[utf8]{inputenc}
\usepackage[T1]{fontenc}
\usepackage{url}
\usepackage{graphicx}
\usepackage{subcaption}
\usepackage{amsmath}
\usepackage{hyperref}

\usepackage{xspace}
\usepackage{soulutf8}
\usepackage{soul}
\usepackage{flushend}
\usepackage{xcolor}

\newcommand{\italico}[1]{\textit{#1}}
\newcommand{\eg}{\italico{e.g.}\xspace}
\newcommand{\ie}{\italico{i.e.}\xspace}

\title{A General-Purpose Diversified 2D Seismic Image Dataset from NAMSS}

\author{
 Lucas de Magalhães Araujo \\
 Instituto de Computação \\
 Universidade Estadual de Campinas (Unicamp) \\
 \texttt{lmaraujo@unicamp.br} \\
    \And
 Otávio Oliveira Napoli \\
 Instituto de Computação \\
 Universidade Estadual de Campinas (Unicamp) \\
 \texttt{otavio.napoli@ic.unicamp.br} \\
    \And
 Sandra Avila \\
 Instituto de Computação \\
 Universidade Estadual de Campinas (Unicamp) \\
 \texttt{sandra@ic.unicamp.br} \\
    \And
 Edson Borin \\
 Instituto de Computação \\
 Universidade Estadual de Campinas (Unicamp) \\
 \texttt{borin@unicamp.br} 
}

\begin{document}

\maketitle

\begin{abstract}
   
We introduce the Unicamp-NAMSS dataset, a large, diverse, and geographically distributed collection of migrated 2D seismic sections designed to support modern machine learning research in geophysics. We constructed the dataset from the National Archive of Marine Seismic Surveys (NAMSS), which contains decades of publicly available marine seismic data acquired across multiple regions, acquisition conditions, and geological settings. After a comprehensive collection and filtering process, we obtained 2\,588 cleaned and standardized seismic sections from 122 survey areas, covering a wide range of vertical and horizontal sampling characteristics. To ensure reliable experimentation, we balanced the dataset so that no survey dominates the distribution, and partitioned it into non-overlapping macro-regions for training, validation, and testing. This region-disjoint split allows robust evaluation of generalization to unseen geological and acquisition conditions. 
We validated the dataset through quantitative and embedding-space analyses using both convolutional and transformer-based models. These analyses showed that Unicamp-NAMSS exhibits substantial variability within and across regions, while maintaining coherent structure across acquisition macro-region and survey types. Comparisons with widely used interpretation datasets (Parihaka and F3 Block) further demonstrated that Unicamp-NAMSS covers a broader portion of the seismic appearance space, making it a strong candidate for machine learning model pretraining. 
The dataset, therefore, provides a valuable resource for machine learning tasks, including self-supervised representation learning, transfer learning, benchmarking supervised tasks such as super-resolution or attribute prediction, and studying domain adaptation in seismic interpretation.

\end{abstract}




\section{Background \& Summary}

Seismic data processing is essential for understanding the subsurface and for tasks such as reservoir characterization, drilling planning, and production optimization. High-quality seismic interpretation reduces uncertainty and supports safer and more efficient decisions in exploration and development.

Deep learning has shown strong potential in seismic applications~\cite{napoli2020accelerating,hecker2023computing,ross2018p,yang2020seismic,wu2019faultseg3d,SEG20-navarro-seismic-attr,harsuko2024optimizing,liu2025shallow}, especially with recent progress in self-supervised learning (SSL) and large-scale pretraining~\cite{oquab2023dinov2,sheng2025seismic}. These methods require large amounts of data, but size alone is not enough. 
Diversity plays a key role in helping models learn robust and general representations~\cite{teterwak2025large}. Recent efforts such as DINOv2~\cite{oquab2023dinov2} highlight this by applying deduplication and data-filtering steps to increase dataset variety and improve model generalization.
Achieving this diversity is challenging in seismic data, where volumes often show strong spatial and temporal correlation, resulting in redundancy and limited variability. 

To address the need for varied and representative data, we introduce the Unicamp-NAMSS dataset, a large collection of 2D seismic images extracted from the National Archive of Marine Seismic Surveys (NAMSS). 
The dataset covers different geographic regions, acquisition periods, environmental conditions, geological settings, and acquisition parameters. This broad variety makes it well suited for SSL and for training larger foundation-style models, since it exposes models to many types of seismic patterns and acquisition scenarios.

As the dataset is unlabeled, it is suited for SSL methods but is also valuable for supervised tasks such as super-resolution~\cite{li2021deep}, denoising~\cite{li2024robust}, and attribute prediction~\cite{napoli2020accelerating,hecker2023computing}, where labels can be generated automatically using standard processing workflows or external tools.
To support these tasks, we provide a predefined train, validation, and test split with no overlap between surveys. 
This design avoids data leakage and ensures that models are evaluated on regions they have never seen before, resulting in a more realistic and reliable assessment of their robustness and generalization.

\section{Methods}


The construction and documentation of the Unicamp-NAMSS dataset follow the methodology proposed by Gebru et al.~in their work ``Datasheets for Datasets''~\cite{Gebru2021}, which outlines principles for dataset creation, including motivation, composition, collection procedures, preprocessing steps, intended uses, and maintenance. Following these guidelines, we organized our data acquisition and documentation to ensure reproducibility, to make explicit the assumptions and design choices involved, and to delimit the scope of validity of subsequent experimental results. This section describes the motivation for assembling a diverse archive of migrated seismic data, the procedures used to locate and extract such data from the NAMSS platform, and the characteristics of the resulting dataset. The complete set of guiding questions proposed by Gebru et al.~is addressed in the Appendix.

A central question at the beginning of this work was whether it would be possible to obtain enough migrated seismic data, with sufficient geographic and acquisition diversity, to build the dataset used in our super-resolution experiments. This task initially motivated this data collection. 
Naturally, this diversity is also beneficial for self-supervised pretraining tasks and other supervised learning applications. 
The search began with the SEG Open Data catalog~\footnote{\url{https://wiki.seg.org/wiki/Open_data}}, which lists publicly available geophysical datasets. 
Although most entries correspond to isolated surveys, one source stood out as particularly suitable: the NAMSS. It contains a large number of historical 2D and 3D surveys from different regions, it offered the volume and variability required for this study.

The NAMSS is maintained by the United States Geological Survey (USGS) as part of its mission to preserve and distribute scientific data and includes surveys acquired directly by USGS, by the Bureau of Ocean Energy Management, and by private companies that choose to contribute their data for research and educational use. 
Due to U.S.~federal regulations, processed seismic data older than 25 years can be publicly released, which explains the concentration of surveys acquired between the mid-1970s and early 1990s. 
All data available on NAMSS are in the public domain and may be freely used for scientific~purposes\footnote{\url{https://www.usgs.gov/information-policies-and-instructions/acknowledging-or-crediting-usgs}}.

\begin{figure}[!hptb]
    \centering
    \includegraphics[width=0.9\textwidth]{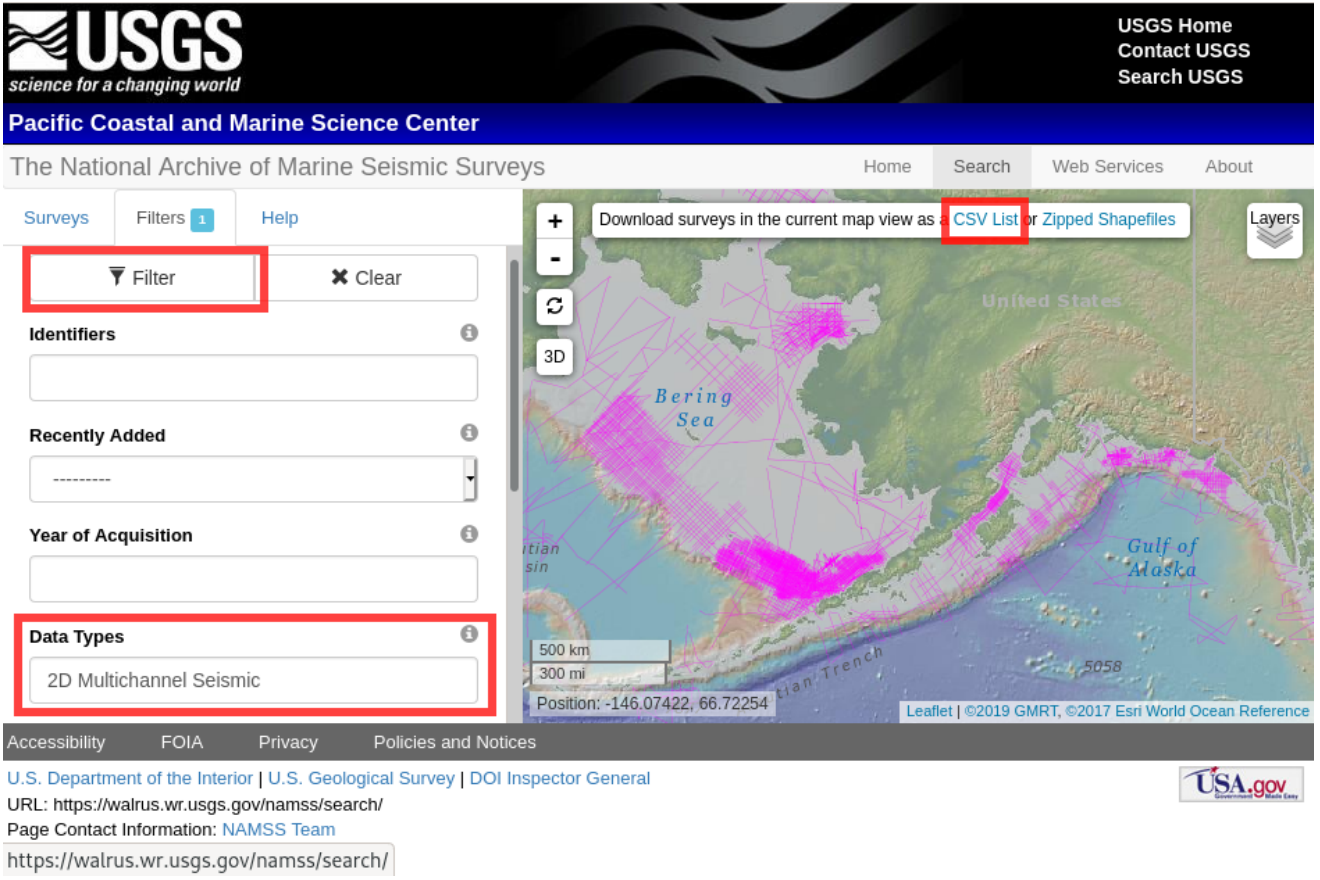}
    \caption{Example of using the NAMSS search tool in the Bering Sea and Gulf of Alaska region. On the left are highlighted examples of search criteria by data type and the filter button, which applies the filter criteria only to data within the visible region. The search result is illustrated on the map as pink lines, representing the seismic lines of each survey. On the right, a link to download a CSV file with information about the surveys displayed on the map is highlighted. Source: \url{https://walrus.wr.usgs.gov/namss/search}.}
    \label{fig:namss_search_tool}
\end{figure}

The archive contained 581 publicly available 2D surveys\footnote{\url{https://walrus.wr.usgs.gov/namss}} to the day we started the data collection.
Data access is provided through an interactive map-based search tool (Figure~\ref{fig:namss_search_tool}), which allows filtering by survey name, acquisition year, and data type. 
However, these filters apply only to the portion of the map currently visible on the screen. Since the interface does not allow zooming out to view all regions simultaneously, it is not possible to apply filters across the entire archive at once. 
Consequently, collecting all relevant migrated 2D surveys required navigating multiple regions manually. 
At the time the dataset was collected, NAMSS offered no unified query API.

Each survey on NAMSS has a dedicated webpage containing detailed metadata, including acquisition parameters (such as source and receiver characteristics), geographic coverage, acquisition dates, processing level, and available file types. 
Surveys can be downloaded either as a single compressed archive or as individual files. 
Typical file types include stacked or migrated seismic data in SEG-Y format (revisions 0 and 1), navigation files, velocity models, reports, and auxiliary figures.

Overall, while NAMSS offers a rich and diverse collection of marine seismic data spanning a wide range of regions and acquisition conditions, its map-based search interface is not optimized for retrieving specific data types on a large scale, such as migrated 2D lines. Addressing these limitations was an essential step in assembling our dataset.





\subsection{Finding 2D migrated data on the NAMSS platform}

We first used the NAMSS map-based tool to locate surveys with 2D multichannel seismic data. By applying the ``2D Seismic Multichannel'' filter across different regions, we compiled a list of 547 distinct 2D multichannel surveys (in addition to 33~singlechannel and 1 sonobuoy surveys, totaling the 581 2D surveys reported on the site).

For each survey in this list, we accessed the XML metadata provided on its webpage and checked whether migrated products were available. Of the 547 multichannel surveys, 143 contained at least one migrated dataset.
All steps following the initial map interaction were automated with scripts. 

For the 143 surveys with migrated products, we then collected the complete inventory of file links and their reported sizes. This resulted in 28\,494 links in total, spanning seismic data files as well as navigation files, velocity models, reports, and figures.

Because per-file metadata beyond size are not consistently provided, we identified migrated seismic files by matching filename patterns with regular expressions. Using this procedure, we marked 9\,350 files as migrated, totaling 157\,GB.



\subsection{Balancing and downloading the dataset}

\begin{figure}[!hptb]
    \centering
    \includegraphics[width=0.8\textwidth]{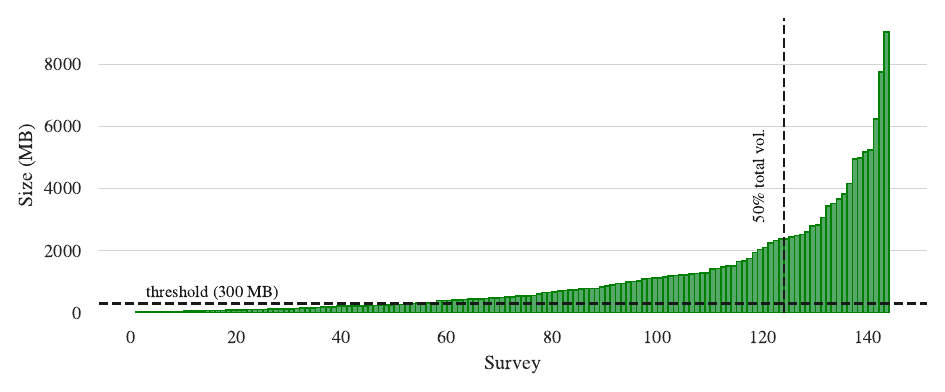}
    \caption{Volume of available migrated data in 143 surveys, ordered by volume. 
    The vertical dashed line shows the point that divides 50\,\% of the total volume. 
    The horizontal dashed line shows the 300\,MB threshold used to truncate the maximum amount of data collected from each survey.
        \label{fig:data_volume}
    }
\end{figure}

After identifying the migrated data available in each survey, we observed a large imbalance in data volume across surveys. As shown in Figure~\ref{fig:data_volume}, the amount of migrated data per survey ranges from 17\,MB to 9\,GB. Among the 143 surveys, half of the total 157\,GB is concentrated in only 18 surveys, while the remaining half is spread across the other 125 surveys.

This imbalance could affect downstream use of the dataset. 
For example, if a user were to train a model for tasks such as super-resolution, denoising, or attribute prediction, surveys with very large volumes of data could dominate the learned representations, reducing the contribution of lower-volume surveys and limiting geographic and geological diversity. 
To reduce this risk, we balanced the dataset so that each survey contributed a similar amount of data.

We defined a target dataset size of approximately 30\,GB, which is comfortably below the total 157\,GB available. 
Based on this target and on the distribution of volumes across surveys, we selected a maximum threshold of 300\,MB per survey, indicated by the horizontal dashed line in Figure~\ref{fig:data_volume}. 
Surveys with less than 300\,MB contributed all available migrated data, while surveys exceeding the threshold were randomly undersampled following the strategy by He and Garcia~\cite{He2009}, which  randomly selects a subset of the available data until the 300\,MB limit is reached.

Using this procedure, the final balanced collection includes at most 300\,MB of migrated data per survey. 
We made all decisions based solely on metadata prior to download. 
After applying the selection rules, we downloaded a total of 3\,210 seismic files, corresponding to 33.6\,GB of migrated seismic data.

\subsection{Preparation, cleaning, and final composition of the set}
\label{ssec:data-preparation}

The downloaded files were in SEG-Y format, which stores seismic samples and metadata in a single file. 
To build a clean and uniform image-based dataset, we extracted only the seismic samples from each trace. 
During this step, values encoded in IBM hexadecimal floating-point format were converted to IEEE 754 double-precision when necessary, and the samples were organized into matrices of size: (number of samples per trace)~$\times$~(number of traces). 
Traces were kept in their original order, and no spatial reordering was applied.

Each matrix was then amplitude-normalized to the interval $[-1,\,1]$ by dividing all samples by the global absolute maximum of the file. 
Formally, if $S$ is the extracted sample matrix, the normalized matrix is
\[
S_{\text{norm}} = \frac{S}{\max_{s \in S} |s|}.
\]
The normalized data were saved in TIFF format, which preserves 4-byte floating-point precision and is convenient for pretraining and downstream processing tasks.

The data-reading software was implemented from the official revision~0 and revision~1 SEG-Y specifications, without the use of external libraries.

After extraction, we performed a cleaning stage. Each file was visually inspected, and 59 corrupted or unreadable files (originating from 7 surveys) were removed. We also detected identical files across multiple surveys, likely due to overlapping survey areas. These duplicates were identified and discarded.

Metadata inspection showed that the collection included a mixture of time- and depth-migrated data with various sampling intervals: 0.1\,ms, 0.5\,ms, 1.0\,ms, 1.6\,ms, 2.0\,ms, 2.5\,ms, 3.0\,ms, 4.0\,ms, and 8.0\,ms for time-migrated data, and 10\,m and 20\,m for depth-migrated data. 
Since 91\,\% of all files had sampling intervals of 4\,ms, we retained only these to maintain uniform vertical sampling across the dataset and removed files with other intervals.

After all preparation and cleaning steps, 622 of the original 3\,210 files were excluded, resulting in a final set of 2\,588~seismic images time-migrated seismic images with 4\,ms sampling interval. 

\section{Data Records}

The instances in the Unicamp-NAMSS dataset were collected from the NAMSS platform, which contains primarily marine seismic surveys (2D and 3D) distributed along the Alaskan coast, the eastern and western coasts of the United States, and in selected regions around Hawaii, Central America, and the Argentine Basin.

All instances are 2D time-migrated seismic sections. In most files, the migration method is not specified; in the minority where this information is available, we observed terms such as Finite Difference Migration, Kirchhoff Time Migration, and True Amplitude Migration. All data share a vertical sampling interval of 4\,ms. 
The average horizontal trace spacing ranges from 4\,m to 305\,m, with 85\,\% of the files lying between 12.5\,m and 50\,m.

\begin{figure}[tb]
    \centering
    \includegraphics[width=0.9\textwidth]{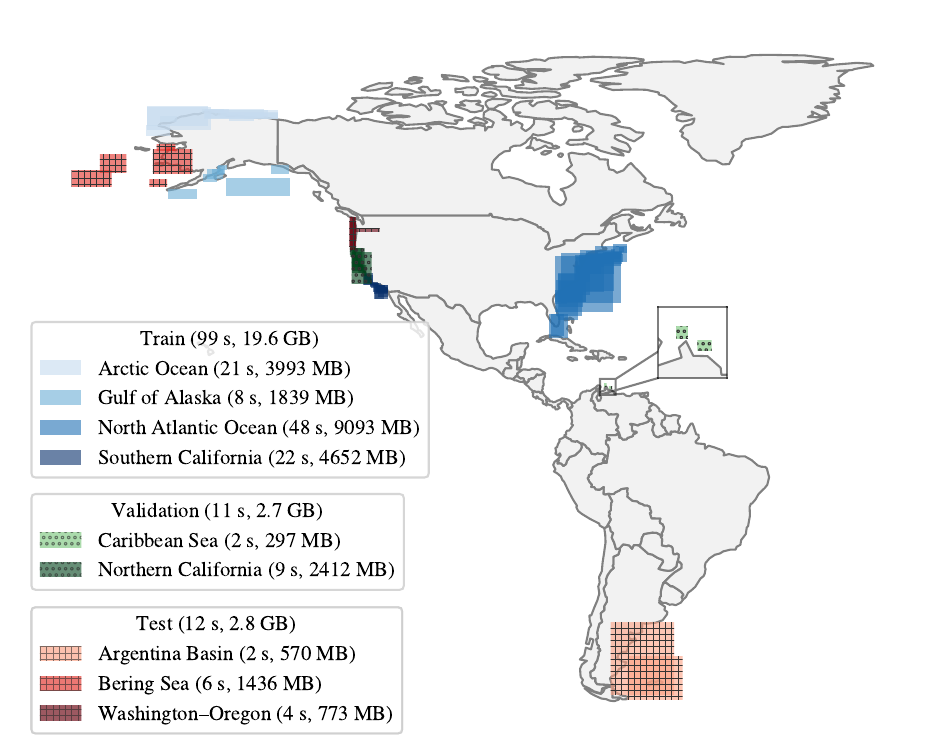}
    \caption{Geographic distribution of the 122 survey areas included in the Unicamp-NAMSS dataset. Each bounding box corresponds to the coverage of a seismic survey. The surveys are divided into Training (blue), Validation (green), and Test (red) subsets, arranged into nine macro-regions with no spatial overlap. The figure also reports, for each macro-region, the number of surveys and the total amount of data it contributes to the dataset.
        \label{fig:coverage_regions}
    }
\end{figure}

In total, the dataset contains 2\,588 seismic sections drawn from 122 distinct survey areas. 
Each section corresponds to a single seismic line (trackline) within a survey. The dataset is balanced so that no survey contributes more than 300\,MB of data. The number of sections per survey ranges from 1 to 73; approximately 70\,\% of surveys contain between 10 and 50 sections. Individual section sizes range from 0.3\,MB to 128\,MB, with 70\,\% falling between 2.5\,MB and 14\,MB. Figure~\ref{fig:coverage_regions} shows the geographic distribution of the survey areas.

The surveys were grouped manually into nine non-overlapping macro-regions to avoid spatial redundancy between regions. These macro-regions were then assigned to training, validation, and test splits as follows:
\begin{itemize}
\itemsep0em
    \item \textbf{Training}: Gulf of Alaska, North Atlantic Ocean, Arctic Ocean, Southern California.
    \item \textbf{Validation}: Caribbean Sea, Northern California.
    \item \textbf{Test}: Argentine Basin, Bering Sea, Washington-Oregon.
\end{itemize}

Because the macro-regions have different data volumes, the split was manually adjusted to achieve approximately 80\,\% training, 10\,\% validation, and 10\,\% testing of the total dataset size. Figure~\ref{fig:coverage_regions} shows the spatial extent of each macro-region, its assignment to a split, and the amount of data contained in each region.

\begin{figure}[tb]
    \centering
    
    \begin{subfigure}[b]{0.48\textwidth}
        \centering
        \includegraphics[width=\textwidth]{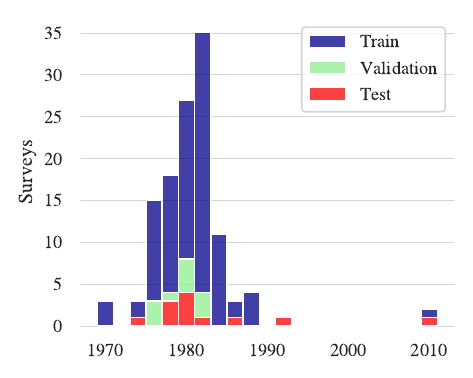}
        \caption{Year}
        \label{fig:surveys_year}
    \end{subfigure}
    \hfill
    \begin{subfigure}[b]{0.48\textwidth}
        \centering
        \includegraphics[width=\textwidth]{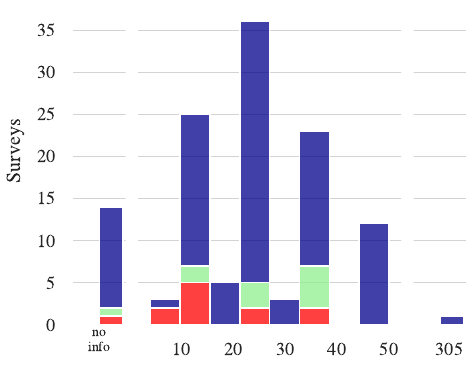}
        \caption{Distance between traces (m)}
        \label{fig:surveys_dx}
    \end{subfigure}

    \vspace{10pt} 

    \begin{subfigure}[b]{0.48\textwidth}
        \centering
        \includegraphics[width=\textwidth]{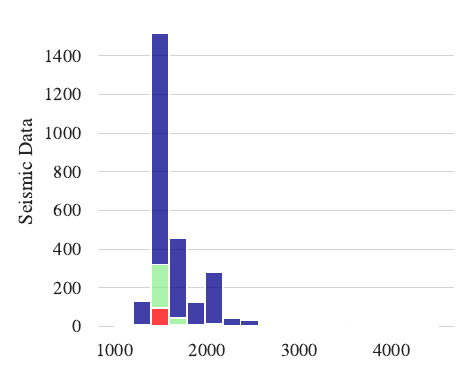}
        \caption{Number of samples per trace}
        \label{fig:samples_trace}
    \end{subfigure}
    \hfill
    \begin{subfigure}[b]{0.48\textwidth}
        \centering
        \includegraphics[width=\textwidth]{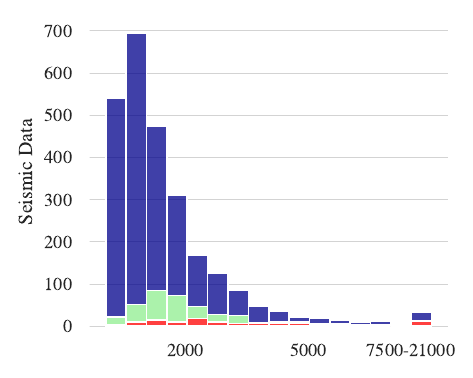}
        \caption{Number of traces per section}
        \label{fig:traces_sample}
    \end{subfigure}
    
    \caption{
        Summary of dataset characteristics for the Training, Validation, and Test subsets. 
        Graphs (a) and (b) describe the 122 surveys, while graphs (c) and (d) summarize the 2\,588 seismic sections. 
        (a) Distribution of survey acquisition years, with 90\,\% conducted between 1975 and 1985. 
        (b) Average trace spacing (dx) per survey; although all data share a fixed temporal sampling rate of 4\,ms, lateral sampling (distance between traces, dx) varies widely, with 12.5\,m, 25\,m, 33\,m (110\,ft), and 50\,m accounting for 70\,\% of the data; 14 surveys lack this information. 
        (c) Number of samples per trace, with 89\,\% of data between 1\,000 and 2\,000 samples or 4\,s to 8\,s. 
        (d) Number of traces per sample, with 80\,\% between 400 and 3\,200 traces.
        \label{fig:set_information}
    }
\end{figure}

Additional dataset characteristics are summarized in Figure~\ref{fig:set_information}. These include: (a) the distribution of survey acquisition years, (b) the average trace spacing for each survey, (c) the number of samples per trace, and (d) the number of traces per section. 
Together, these figures highlight the diversity of acquisition periods, spatial sampling, and image dimensions represented in the~dataset.


In summary, the Unicamp-NAMSS dataset provides a diverse, balanced, and geographically disjoint collection of 2D migrated seismic sections. 
Its variety in acquisition parameters, survey locations, and image dimensions makes it suitable for a wide range of machine learning tasks, including self-supervised pretraining and other seismic processing tasks.

Although the dataset has been carefully cleaned and standardized, we note that data quality varies across surveys due to differences in acquisition age, instrumentation, and processing workflows. 
For applications that require high-quality or artifact-free data, expert inspection of selected instances may be advisable. 
The dataset is intended primarily for research, benchmarking, and method development, rather than for direct deployment in production workflows.

\section{Technical Validation}

To evaluate the quality, diversity, and representativeness of the Unicamp-NAMSS dataset, we conducted analyses to characterize variability both within and across geographic regions.

\subsection{Embedding-space variability}

To analyze diversity at a feature level, we projected the dataset into a learned embedding space. 
The samples are first normalized using z-normalization per image (\ie, per section), then passed through a pretrained model to extract feature embeddings. 
The high-dimensional embeddings are then projected to two dimensions using Uniform Manifold Approximation and Projection (UMAP)~\cite{mcinnes2018umap} for visualization. 
This approach allows us to visualize structural and textural variability in the dataset, as well as to examine potential correlations with metadata such as acquisition macro-region and dataset split.

We extracted embeddings using two well-known models for computer vision: (i) a ResNet-50 pretrained on COCO~\cite{lin2014microsoft}, and (ii) a Vision Transformer (ViT-B/14) pretrained with DINOv2~\cite{oquab2023dinov2} on the LVD-142M dataset. 
These models were chosen because they provide broad, general visual representations that allow us to examine structural and textural variability in the seismic~images.

\begin{figure}[p]
    \centering

    \begin{minipage}[b]{0.50\textwidth}
        \centering
        \includegraphics[width=\textwidth]{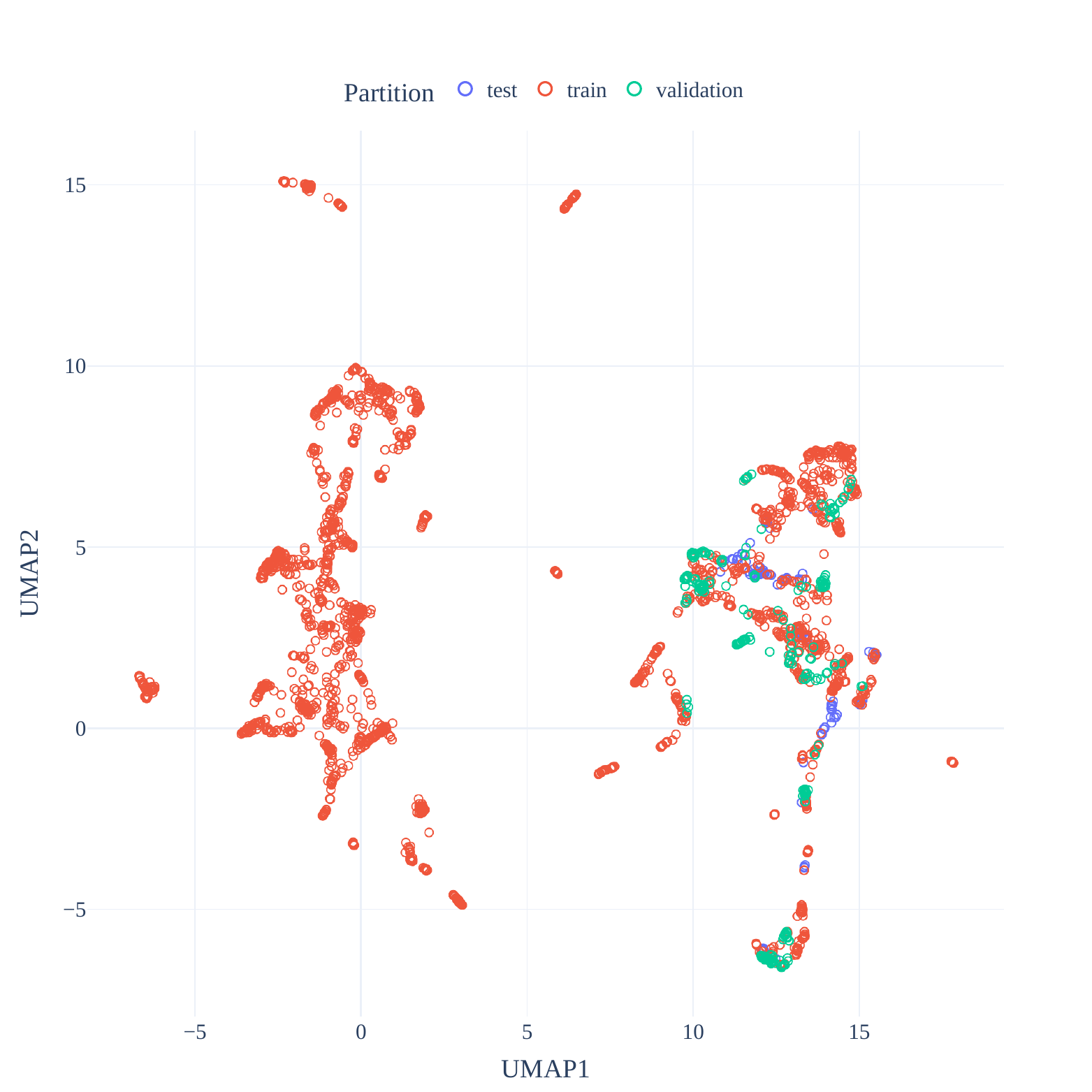}
        \par\medskip
        \textbf{(a)} ResNet-50, colored by dataset split
    \end{minipage}\hfill
    \begin{minipage}[b]{0.50\textwidth}
        \centering
        \includegraphics[width=\textwidth]{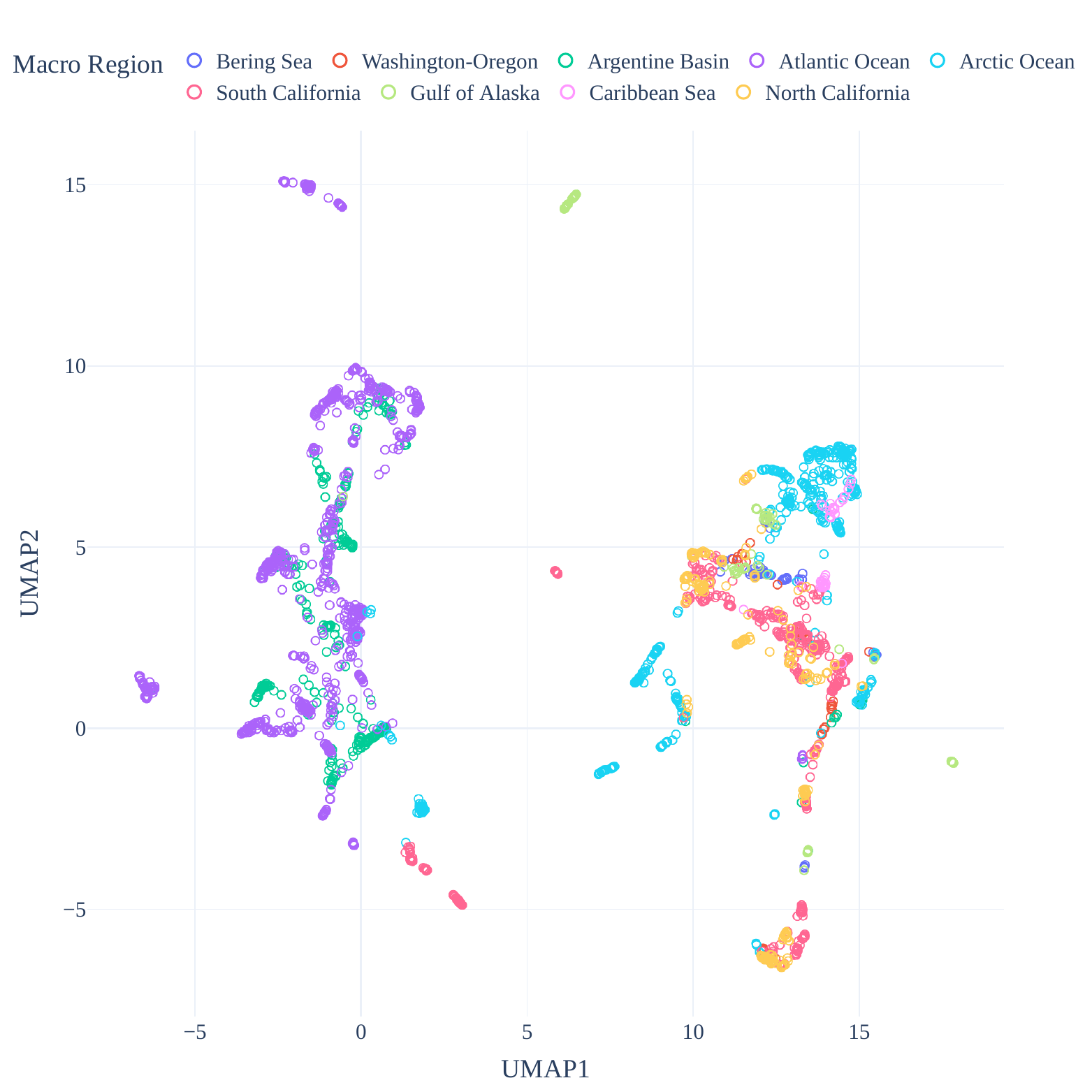}
        \par\medskip
        \textbf{(b)} ResNet-50, colored by acquisition macro-region
    \end{minipage}

    \vspace{1em}

    \begin{minipage}[b]{0.50\textwidth}
        \centering
        \includegraphics[width=\textwidth]{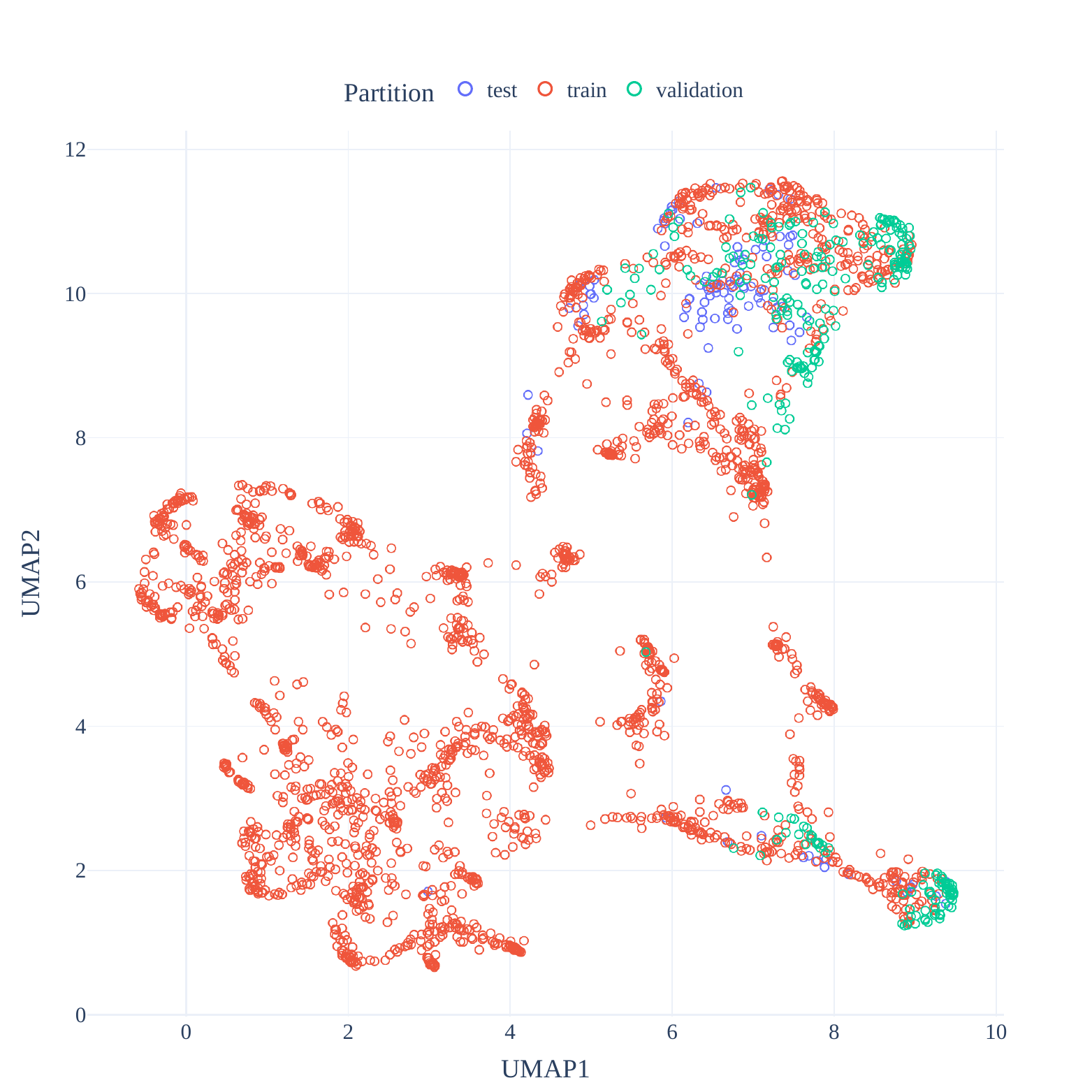}
        \par\medskip
        \textbf{(c)} DINOv2 ViT-B/14, colored by dataset split
    \end{minipage}\hfill
    \begin{minipage}[b]{0.50\textwidth}
        \centering
        \includegraphics[width=\textwidth]{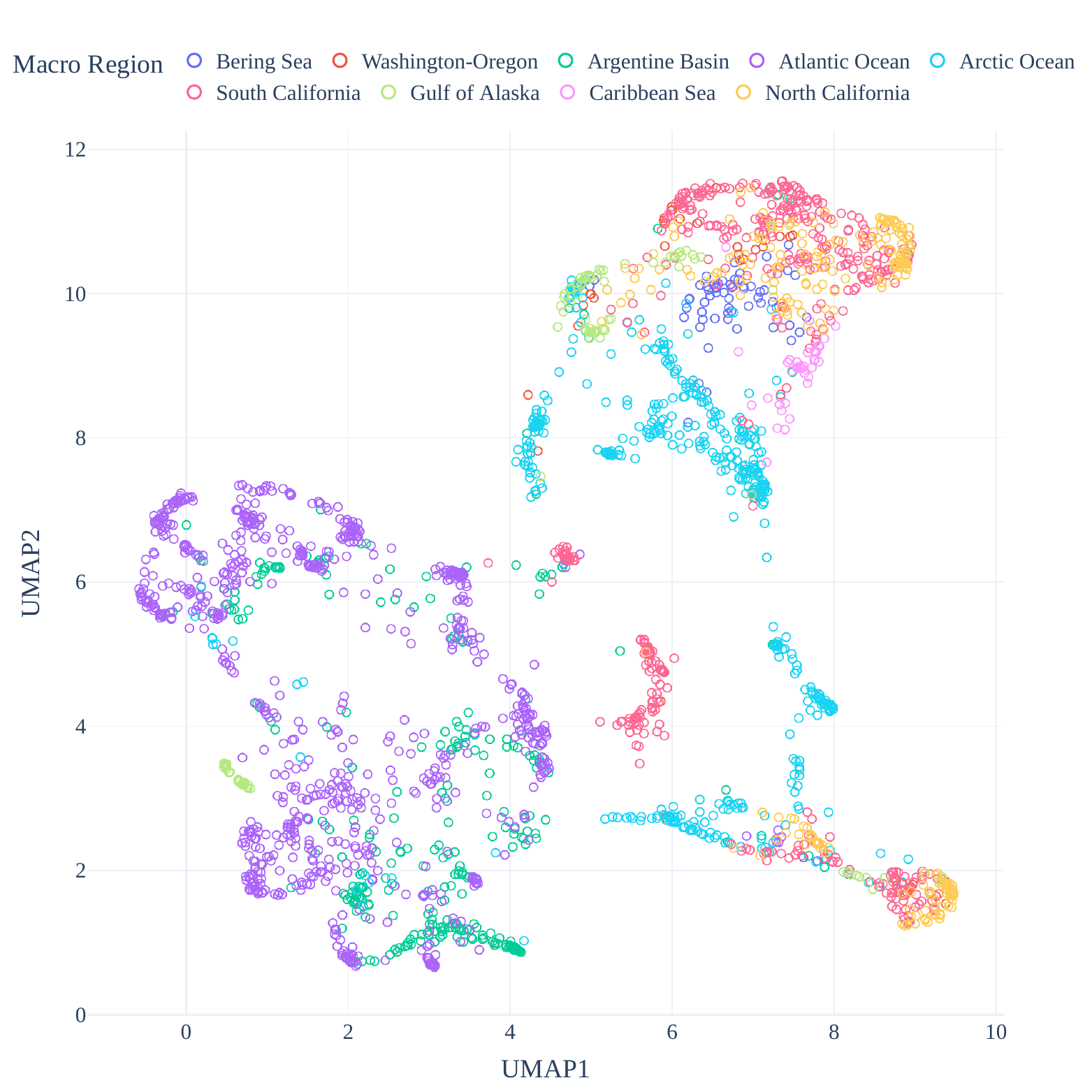}
        \par\medskip
        \textbf{(d)} DINOv2 ViT-B/14, colored by macro-region
    \end{minipage}

    \caption{UMAP projections of NAMSS embeddings extracted using two pretrained models.  
        Top row: ResNet-50 pretrained on COCO.  
        Bottom row: DINOv2 ViT-B/14.  
        Left column: samples colored by dataset split.  
        Right column: samples colored by acquisition macro-region.  
        Training data span a wide region of embedding space, while validation and test samples appear more concentrated. No clear clustering by acquisition macro-region is observed.
        \label{fig:resnet_dinov2_umap}
    }
\end{figure}

Figures~\ref{fig:resnet_dinov2_umap}\,(a) and (c) show UMAP projections colored by dataset split. 
In both models, the training data occupy a wide region of embedding space, indicating a high degree of variability. 
Validation and test samples appear in more compact clusters, suggesting greater internal homogeneity. 
Importantly, the training distribution encompasses the validation and test regions, meaning that the training set covers the range of structures present in the other splits, a desirable characteristic for evaluating generalization performance.

Acquisition location can affect visual appearance through variations in acquisition equipment and processing workflows. 
In practice, each region may be interpreted as its own domain.
Figures~\ref{fig:resnet_dinov2_umap}\,(b) and (d) present the same embeddings, now colored by acquisition macro-region. 
A clear separation is visible in the DINOv2 embeddings (Figure~\ref{fig:resnet_dinov2_umap}\,(d)), particularly for surveys from the Bering Sea and Arctic Ocean (both from train splits). In contrast, this pattern is far less pronounced in the ResNet-50 embeddings (Figure~\ref{fig:resnet_dinov2_umap}\,(b)), where samples from different macro-regions are more intermingled.

These observations suggest that no single geographic region dominates the dataset and that samples from different macro-regions are broadly dispersed across the embedding space. Instead, geological setting, survey geometry, and processing workflows appear to be the primary sources of structural variability. Such diversity is advantageous for representation learning, as it exposes models to a richer set of seismic patterns than would be captured by the acquisition period alone.


\subsection{Relationship of Unicamp-NAMSS to other seismic datasets}

\begin{figure}[!hptb]
    \centering

    \begin{minipage}[b]{0.98\textwidth}
        \centering
        \includegraphics[width=\textwidth]{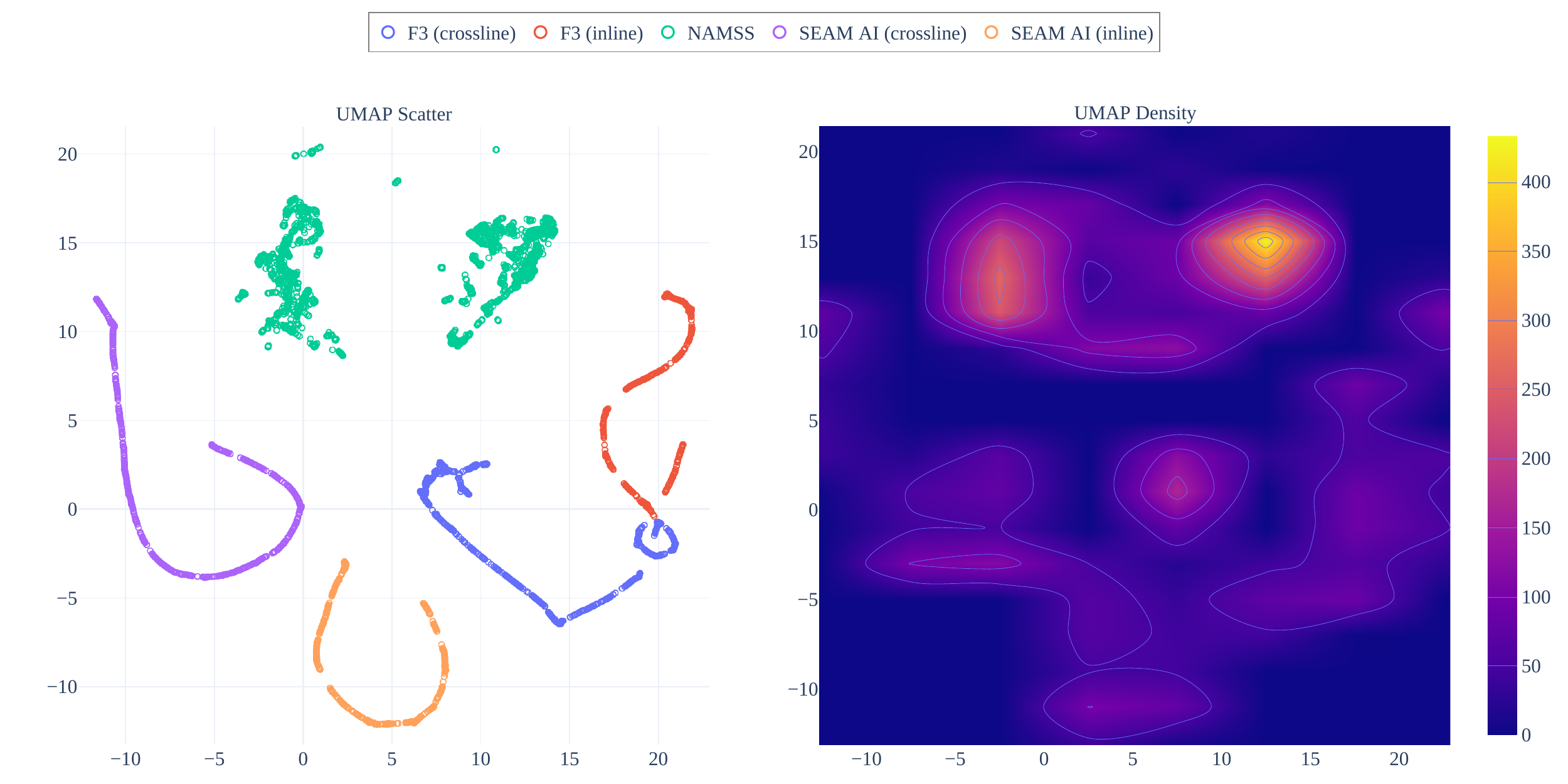}
        \par\medskip
        \textbf{(a)} ResNet-50 (COCO-pretrained)
    \end{minipage}

    \begin{minipage}[b]{0.98\textwidth}
        \centering
        \includegraphics[width=\textwidth]{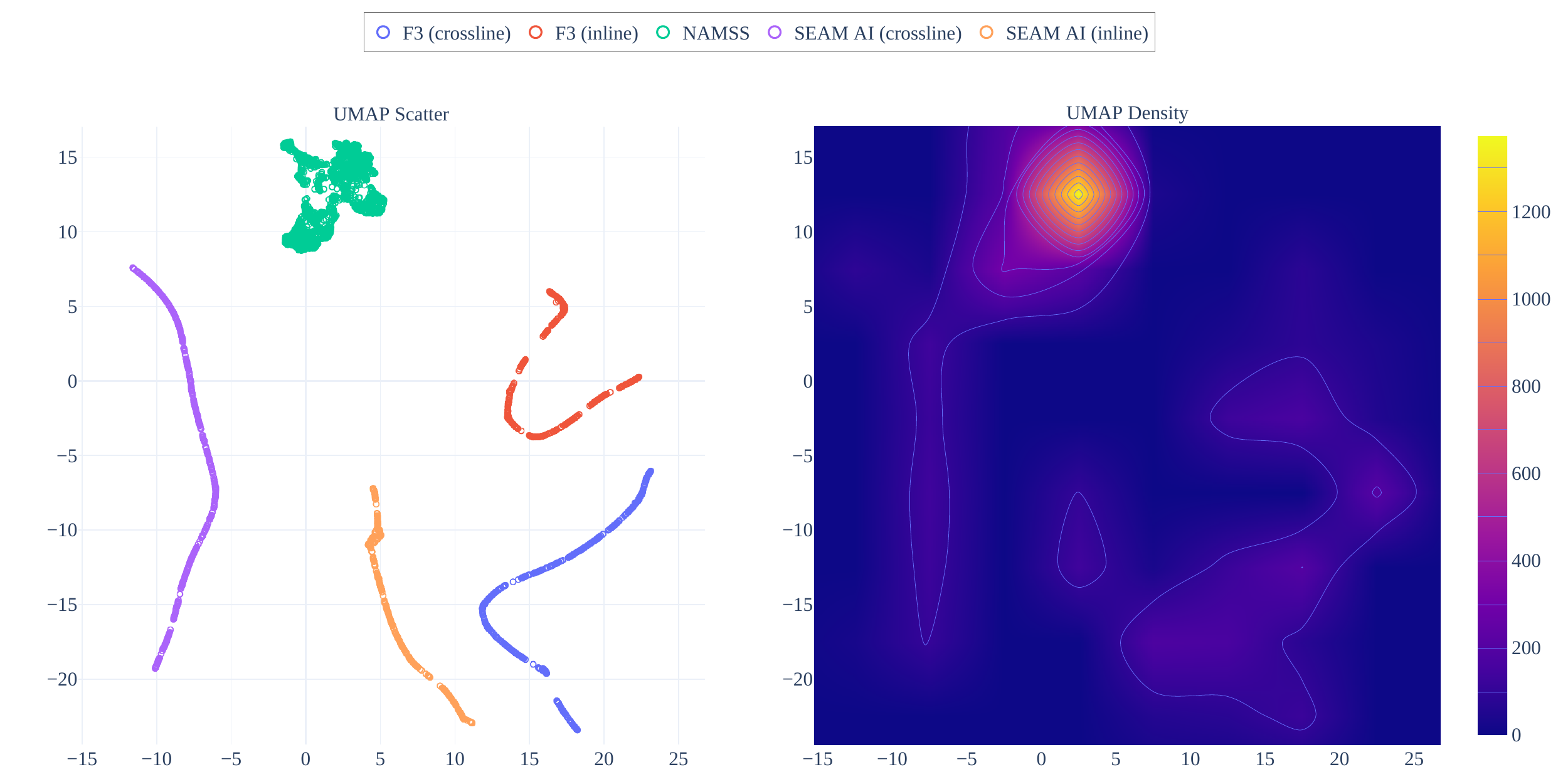}
        \par\medskip
        \textbf{(b)} DINOv2 ViT-B/14
    \end{minipage}

    \caption{UMAP visualizations comparing Unicamp-NAMSS with the Parihaka and Netherlands F3 Block datasets using embeddings extracted from (a) ResNet-50 and (b) DINOv2 ViT-B/14. The 3D volumes were sampled along both inlines and crosslines. Left panels show scatter plots, while right panels show density plots highlighting sample concentration. Unicamp-NAMSS spans a much broader region of the embedding space, reflecting greater diversity in geological structures and acquisition conditions, as compared to the more localized and homogeneous Parihaka and F3 datasets.}
    \label{fig:datasets_resnet_dinov2_umap}
\end{figure}

To understand how Unicamp-NAMSS compares to commonly used seismic interpretation datasets, we compared it with two widely adopted benchmarks: the Parihaka dataset~\cite{aicrowd_seamai} (New Zealand) and the Netherlands F3 Block~\cite{alaudah2019machine} (North Sea). As these datasets are 3D seismic volumes, we extracted both inlines and crosslines to obtain 2D sections comparable to the Unicamp-NAMSS images. All datasets were normalized using per-image z-normalization before feature extraction using ResNet-50 and DINOv2 ViT-B/14.
The resulting UMAP projections are shown in Figure~\ref{fig:datasets_resnet_dinov2_umap}, where left panels present scatter plots and right panels show density maps from the same UMAP coordinates.

Unicamp-NAMSS occupies a broad region of the embedding space in both models (panels on left), reflecting its geographic and acquisition diversity. 
In contrast, Parihaka and F3 form tight, well-separated clusters, due to their origin from single, geographically localized surveys with more uniform acquisition parameters and geological settings. 
Within each of these datasets, neighbor lines (inlines or crosslines) cluster closely and often align along smooth trajectories in latent space, producing compact structures not observed in the more heterogeneous Unicamp-NAMSS distribution. The fact that Parihaka and F3 do not overlap with each other, nor between their inline and crossline subsets, further indicates that directionality and local geology both shape their internal variability.

Density plots reveal additional structure: certain regions of the Unicamp-NAMSS embedding space exhibit high sample concentration, suggesting the presence of geological patterns that recur across multiple surveys. However, these dense regions remain distinct from the clusters formed by Parihaka and F3, implying that they contain characteristic features not commonly present in Unicamp-NAMSS.

Overall, the clear separation between Unicamp-NAMSS and the two benchmark datasets highlights two important implications. First, the wide coverage of the Unicamp-NAMSS embedding space suggests that it may serve as a strong foundation for pretraining, offering broad exposure to diverse acquisition settings and geological contexts. 
Second, dataset-specific fine-tuning or domain adaptation may still be advisable when transferring models to Parihaka, F3, or other specialized domains.

\section{Code and Data Availability}

All codes are open-source, licensed under MIT License, and available at \url{https://github.com/discovery-unicamp/namss-dataset}. The dataset is available at \url{https://zenodo.org/records/18330487}.

\section*{Acknowledgements}

The authors acknowledge the Discovery Lab for providing the experimental infrastructure and the National Archive of Marine Seismic Surveys (NAMSS) for making the data available. S.~Avila is partially funded by CNPq (grant 316489/2023-9), and FAPESP (grant 2023/12086-9, 2023/12865-8, 2020/09838-0, 2013/08293-7). E.~Borin acknowledges support from CNPq (grant 315399/2023-6) and FAPESP (grant 2013/08293-7).

\section*{Author contributions statement}

The dataset was created by Lucas de Magalhães Araujo, Edson Borin, and Sandra Avila, respectively a master's student and faculty members at the Institute of Computing, University of Campinas (Unicamp). Otávio Napoli, a PhD student at the same institution, contributed to data analysis, technical validation, and writing.

\section*{Competing interests} 

The authors declare no competing interest, financial and non-financial.


\newpage
\bibliographystyle{unsrt}
\bibliography{bibliography}

@article{Gebru2021,
  author  = {Timnit Gebru and Jamie Morgenstern and Briana Vecchione and Jennifer Wortman Vaughan and Hanna Wallach and Hal Daumé III and Kate Crawford},
  title   = {Datasheets for Datasets},
  journal = {Communications of the ACM},
  volume  = {64},
  number  = {12},
  pages   = {86--92},
  year    = {2021},
  month   = {November},
  issn    = {0001-0782},
}

@article{He2009,
  author  = {Haibo He and Edwardo A. Garcia},
  title   = {Learning from Imbalanced Data},
  journal = {IEEE Transactions on Knowledge and Data Engineering},
  volume  = {21},
  number  = {9},
  pages   = {1263--1284},
  year    = {2009},
}

@article{napoli2020accelerating,
  title   = {Accelerating multi-attribute unsupervised seismic facies analysis with rapids},
  author  = {Napoli, Ot{\'a}vio O and Rosario, Vanderson Martins do and Navarro, Jo{\~a}o Paulo and Borin, Edson and others},
  journal = {arXiv preprint arXiv:2007.15152},
  year    = {2020}
}

@article{oquab2023dinov2,
  title   = {{DINOv2: L}earning robust visual features without supervision},
  author  = {Oquab, Maxime and Darcet, Timoth{\'e}e and Moutakanni, Th{\'e}o and Vo, Huy and Szafraniec, Marc and Khalidov, Vasil and Fernandez, Pierre and Haziza, Daniel and Massa, Francisco and El-Nouby, Alaaeldin and others},
  journal = {Transactions on Machine Learning Research},
  year    = {2024}
}

@article{alaudah2019machine,
  title     = {A machine-learning benchmark for facies classification},
  author    = {Alaudah, Yazeed and Micha{\l}owicz, Patrycja and Alfarraj, Motaz and AlRegib, Ghassan},
  journal   = {Interpretation},
  volume    = {7},
  number    = {3},
  pages     = {SE175--SE187},
  year      = {2019},
  publisher = {Society of Exploration Geophysicists and American Association of Petroleum~…}
}

@misc{aicrowd_seamai,
  author       = {SEAM AI},
  title        = {Seismic Facies Identification Challenge},
  howpublished = {\url{https://www.aicrowd.com/challenges/seismic-facies-identification-challenge}},
  note         = {Accessed: 2025-11-07},
  year         = {2021}
}

@inproceedings{hecker2023computing,
  title        = {Computing seismic attributes with deep-learning models},
  author       = {Hecker, N{\'\i}colas and Napoli, Ot{\'a}vio O and Astudillo, Carlos A and Navarro, Jo{\~a}o Paulo and Souza, Alan and Miranda, Daniel and Villas, Leandro A and Borin, Edson},
  booktitle    = {2023 International Symposium on Computer Architecture and High Performance Computing Workshops (SBAC-PADW)},
  pages        = {31--35},
  year         = {2023},
  organization = {IEEE}
}

@article{ross2018p,
  title     = {P wave arrival picking and first-motion polarity determination with deep learning},
  author    = {Ross, Zachary E and Meier, Men-Andrin and Hauksson, Egill},
  journal   = {Journal of Geophysical Research: Solid Earth},
  volume    = {123},
  number    = {6},
  pages     = {5120--5129},
  year      = {2018},
  publisher = {Wiley Online Library}
}

@article{yang2020seismic,
  title     = {Seismic horizon tracking using a deep convolutional neural network},
  author    = {Yang, Liuxin and Sun, Sam Zandong},
  journal   = {Journal of Petroleum Science and Engineering},
  volume    = {187},
  pages     = {106709},
  year      = {2020},
  publisher = {Elsevier}
}

@article{wu2019faultseg3d,
  title     = {FaultSeg3D: Using synthetic data sets to train an end-to-end convolutional neural network for 3D seismic fault segmentation},
  author    = {Wu, Xinming and Liang, Luming and Shi, Yunzhi and Fomel, Sergey},
  journal   = {Geophysics},
  volume    = {84},
  number    = {3},
  pages     = {IM35--IM45},
  year      = {2019},
  publisher = {Society of Exploration Geophysicists}
}

@inbook{SEG20-navarro-seismic-attr,
  author    = {João Paulo Navarro and Pedro Mário Cruz e Silva and Doris Pan and Ken Hester},
  title     = {Real-time seismic attributes computation with conditional GANs},
  booktitle = {SEG Technical Program Expanded Abstracts 2020},
  chapter   = {},
  pages     = {1611-1615},
  year      = {2020},
  doi       = {10.1190/segam2020-3427757.1},
  url       = {https://library.seg.org/doi/abs/10.1190/segam2020-3427757.1},
  publisher = {Society of Exploration Geophysicists}
}

@article{sheng2025seismic,
  title     = {Seismic foundation model: A next generation deep-learning model in geophysics},
  author    = {Sheng, Hanlin and Wu, Xinming and Si, Xu and Li, Jintao and Zhang, Sibo and Duan, Xudong},
  journal   = {Geophysics},
  volume    = {90},
  number    = {2},
  pages     = {IM59--IM79},
  year      = {2025},
  publisher = {Society of Exploration Geophysicists}
}

@article{harsuko2024optimizing,
  title     = {Optimizing a transformer-based network for a deep-learning seismic processing workflow},
  author    = {Harsuko, Randy and Alkhalifah, Tariq},
  journal   = {Geophysics},
  volume    = {89},
  number    = {4},
  pages     = {V347--V359},
  year      = {2024},
  publisher = {Society of Exploration Geophysicists}
}

@article{liu2025shallow,
  title     = {From shallow to deep: Enhancing seismic resolution with weak supervision},
  author    = {Liu, Dawei and He, Yijie and Wang, Xiaokai and Sacchi, Mauricio D and Chen, Wenchao and Du, Guanghong and Zhang, Mengbo},
  journal   = {Geophysics},
  volume    = {90},
  number    = {3},
  pages     = {V223--V239},
  year      = {2025},
  publisher = {Society of Exploration Geophysicists}
}

@inproceedings{teterwak2025large,
  title     = {Is Large-scale Pretraining the Secret to Good Domain Generalization?},
  author    = {Teterwak, Piotr and Saito, Kuniaki and Tsiligkaridis, Theodoros and Plummer, Bryan A and Saenko, Kate},
  booktitle = {International Conference on Learning Representations},
  year      = {2025}
}

@inproceedings{lin2014microsoft,
  title        = {{Microsoft COCO: C}ommon objects in context},
  author       = {Lin, Tsung-Yi and Maire, Michael and Belongie, Serge and Hays, James and Perona, Pietro and Ramanan, Deva and Doll{\'a}r, Piotr and Zitnick, C Lawrence},
  booktitle    = {European Conference on Computer Vision},
  pages        = {740--755},
  year         = {2014},
  organization = {Springer}
}

@article{li2021deep,
  title     = {Deep learning for simultaneous seismic image super-resolution and denoising},
  author    = {Li, Jintao and Wu, Xinming and Hu, Zhanxuan},
  journal   = {IEEE Transactions on Geoscience and Remote Sensing},
  volume    = {60},
  pages     = {1--11},
  year      = {2021},
  publisher = {IEEE}
}

@article{li2024robust,
  title     = {Robust seismic data denoising via self-supervised deep learning},
  author    = {Li, Ji and Trad, Daniel and Liu, Dawei},
  journal   = {Geophysics},
  volume    = {89},
  number    = {5},
  pages     = {V437--V451},
  year      = {2024},
  publisher = {Society of Exploration Geophysicists}
}

@article{mcinnes2018umap,
  title   = {{UMAP: U}niform manifold approximation and projection for dimension reduction},
  author  = {McInnes, Leland and Healy, John and Melville, James},
  journal = {Journal of Open Source Software},
  
  year    = {2018}
}

\newpage
\appendix

\section{Appendix: Datasheet for the Unicamp-NAMSS 2D Seismic Image Dataset}
\label{appendix:datasheet}

This datasheet follows the structure proposed by Gebru et al.~\cite{Gebru2021} in the paper ``Datasheets for Datasets''. Only the questions that are relevant to this dataset are answered. The dataset described here consists of 2D migrated seismic images extracted from the National Archive of Marine Seismic Surveys (NAMSS). The objective of this documentation is to provide transparency regarding the dataset's motivation, composition, collection process, preprocessing steps, recommended uses, and limitations.

\subsection{Motivation}

\paragraph{For what purpose was the dataset created? Was there a specific task in mind? Was there a gap that needed to be filled?\\}

The dataset was created to support research in machine learning for seismic data, enabling tasks such as super-resolution, self-supervised representation learning, and other data-driven seismic analysis methods. 
At the time of its construction, no publicly available seismic dataset offered the level of diversity in migrated 2D seismic images required for large-scale comparative evaluations of deep-learning models. 
Most existing datasets originate from a single exploration area and therefore lack the variability needed to assess model robustness, often resulting in biased conclusions. 
In contrast, the NAMSS archive contains surveys acquired in geographically distinct regions and under varied acquisition conditions, providing the diversity necessary to construct a more representative and comprehensive dataset.

\paragraph{Who created the dataset?\\}

The dataset was created by Lucas de Magalhães Araujo, Edson Borin, and Sandra Avila, respectively a master's student and faculty members at the Institute of Computing, University of Campinas (Unicamp). Otávio Napoli, a PhD student at the same institution, contributed to data analysis, technical validation, and writing.

\paragraph{Who funded the creation of the dataset?\\}

The dataset did not receive direct funding for its creation.

\subsection{Composition}

\paragraph{What do the instances in the dataset represent?\\}

Each instance is a single 2D migrated seismic line obtained from the National Archive of Marine Seismic Surveys (NAMSS). The temporal sampling interval of the data is 4\,ms, while the spatial trace spacing ranges from 4\,m to 305\,m, with approximately 85\,\% of the lines between 12.5\,m and 50\,m. 

\paragraph{How many instances are in the dataset?\\}

The dataset contains 2\,588 seismic images from 122 exploration areas. Approximately 70\,\% of the images range from 2.5\,MB to 14\,MB in size, with a median of 7\,MB, a minimum of 0.3\,MB, and a maximum of 128\,MB.

\paragraph{Does the dataset contain all possible instances, or is it a sample of a larger set? If it is a sample, is it representative?\\}

At the time of the collection (2021--2022), the NAMSS platform consisted primarily of 2D and 3D marine seismic data from regions near Alaska, the U.S.~East and West Coasts, the Caribbean, Hawaii, Central America, and the Argentine Basin. We identified 9\,350 migrated 2D seismic files from 143 exploration areas. From these, 3\,210 files were randomly selected, with a cap of 300 MB per area to maintain balance. After cleaning, 2\,588 files from 122 areas remained.

Considering only the NAMSS archive, this selection represents approximately 28\,\% of all available migrated 2D data, sampled in a geographically balanced manner across the entire archive. However, NAMSS itself is not representative of the full spectrum of possible seismic imaging modalities or subsurface settings worldwide.

\paragraph{What data does each instance consist of?\\}

Each instance is a image derived from a SEG-Y seismic file. All seismic samples were extracted from their original traces and converted from IBM hexadecimal floating-point format to IEEE floating-point format when required. The samples were arranged into a matrix of size (number of samples per trace $\times$ number of traces), normalized to the interval $[-1,\,1]$, and saved in TIFF format.

\paragraph{Are there labels or targets associated with the data?\\}

No.

\paragraph{Are any instances missing information?\\}

No instances are missing required information.

\paragraph{Are there relationships between instances?\\}

Yes. Instances are grouped according to their exploration area, following the NAMSS structure. Multiple seismic lines may originate from the same survey area.

\paragraph{Are there recommended train, validation, and test splits?\\}

Yes. The data were manually divided into nine macroregions without geographic overlap between them. These macroregions were then assigned to dataset splits as follows:

\begin{itemize}
\itemsep0em
\item \textbf{Training}: Gulf of Alaska, North Atlantic Ocean, Arctic Ocean, Southern California.
\item \textbf{Validation}: Caribbean Sea, Northern California.
\item \textbf{Test}: Argentine Basin, Bering Sea, Washington-Oregon.
\end{itemize}

Because the macroregions differ significantly in volume, the split was manually adjusted to yield approximately 80\,\% training, 10\,\% validation, and 10\,\% test. Each subset is stored in separate directories.

\paragraph{Are there errors, noise, or redundancies?\\}

Duplicate files were removed. Variations in image quality, noise levels, blurring, and processing artifacts exist across the dataset. Additionally, corrupted files were removed during the cleaning process.

\paragraph{Is the dataset self-contained?\\}

Yes. All processed images and their corresponding raw SEG-Y files are included.

\subsection{Collection Process}

\paragraph{How were the data collected? What mechanisms or procedures were used?\\}

To obtain a dataset suitable for self-supervised learning, we looked for a large and diverse collection of 2D migrated seismic images. The NAMSS platform met these requirements, but it does not offer a global search interface. Instead, filters apply only to the region currently visible on the map, and it is not possible to determine from the global interface which surveys include migrated data. Therefore, we executed the following procedure:

\begin{enumerate}
\itemsep0em
\item Using the NAMSS search tool, we identified all exploration areas with 2D seismic data by applying the ``2D Seismic Multichannel'' filter across different map regions. This produced a list of 547 distinct 2D multichannel surveys, each with a link to its detail page.
\item For each survey page, we inspected the available XML metadata to determine whether migrated data were present. Out of the initial 547 surveys, 143 contained migrated data.
\item For these 143 surveys, we scraped the complete inventory of available files, including file sizes. This resulted in a total of 28,494 files, encompassing seismic data, navigation data, velocity models, figures, documentation, and reports.
\item Beyond file sizes, no additional metadata were provided. Migrated seismic files were identified by matching filename patterns using regular expressions. Through this process, we identified 9,350 migrated files (157\,GB total).
\item To maintain a manageable dataset size and ensure balance across surveys, we randomly sampled up to 300 MB per survey, yielding 3\,210 links totaling 33.6\,GB.
\item The selected files were downloaded, completing the collection stage. Additional cleaning steps are described in the next section.
\end{enumerate}

Except for Step 1, all steps were automated using Python and Bash scripts.

\paragraph{How was the collection process validated?\\}

The NAMSS homepage at the time listed 581 available 2D surveys. Our procedure recovered 547 multichannel surveys, 33 single-channel surveys, and one sonobuoy survey, matching the expected total.  
Step 4, which relied on filename patterns to determine whether a file contained migrated data, was validated through sampling. Although potential misclassification cannot be completely factored out, the verification suggests that such errors are minimal.

\paragraph{From where and when were the data originally obtained?\\}

The original seismic surveys were conducted by various government agencies, such as the USGS, or by private exploration companies (\eg, Exxon, Mobil, Texaco). Metadata available in NAMSS contain limited information about the acquisition, processing, and migration methods.

\paragraph{Over what time period were the data collected?\\}

The original seismic data were acquired between 1969 and 2011, with approximately 90\,\% acquired between 1975 and 1985. Metadata collection and data download from NAMSS were performed in April 2021.

\paragraph{What are the copyright or licensing restrictions?\\}

According to the U.S.~Department of the Interior, materials produced by federal agencies are in the public domain and may be freely reproduced. See: \url{https://www.doi.gov/copyright}.

\subsection{Preprocessing, Cleaning, and Labeling}

\paragraph{Was any preprocessing performed?\\}

SEG-Y files interleave metadata and seismic traces. To convert them into images, we applied the following steps:

\begin{itemize}
\itemsep0em
\item Reading: seismic samples were extracted from each trace and converted from IBM hexadecimal floating-point format to IEEE format when necessary. The traces were assembled into a matrix (samples per trace $\times$ number of traces), maintaining their original order.
\item Normalization: each image was scaled to the interval $[-1,\,1]$ by dividing all samples by the global absolute maximum.
\item Writing: each image was saved in TIFF format.
\end{itemize}

The software for reading SEG-Y files was implemented from the official specification manuals (Revisions 0 and 1), without external libraries.

\paragraph{Was any cleaning performed?\\}

Yes. Duplicate exploration areas were removed (\eg, survey W-62-77-AR was fully contained within B-15-77-AK2).  
To maintain consistent vertical resolution across the dataset, only data with a sampling interval of 4 ms were retained, which represented 91\,\% of all collected files.  
A visual inspection identified 51 corrupted files from seven exploration areas; these were discarded.  
After all cleaning steps, 622 of the 3\,210 initial files were removed, resulting in 2\,588 final instances.

\paragraph{Were the data labeled?\\}

No.

\paragraph{Were the raw data saved?\\}

Yes. All raw SEG-Y files corresponding to the processed images are included.

\paragraph{Are preprocessing or cleaning code available?\\}

Yes. Preprocessing and analysis code are available at \url{https://github.com/discovery-unicamp/namss-dataset}.

\subsection{Uses}

\paragraph{Has the dataset been used for any tasks?\\}

No.

\paragraph{Is there a repository linking publications or systems that use the dataset?\\}

No.

\paragraph{What other tasks could the dataset be used for?\\}

The dataset's strengths lie in its large variability, balanced sampling across surveys, and clean separation of training, validation, and test regions. These characteristics make it well suited for self-supervised representation learning as well as tasks in which labels can be automatically derived from the data, such as denoising, reconstruction, super-resolution, and related seismic processing or geophysical machine-learning applications.

\paragraph{Are there tasks for which the dataset should not be used?\\}

The dataset exhibits variable image quality and may contain processing artifacts. For applications that require high-fidelity seismic imaging or geological interpretation, a domain expert should assess data quality before use.  
The dataset was created primarily for evaluating deep-learning models rather than for deployment-quality training. A possible workflow is to use this dataset for model selection (\eg, comparing architectures), and then perform final training using high-quality proprietary seismic datasets when available.

\end{document}